\documentclass[twoside]{article}
\usepackage{fleqn,espcrc2}
\usepackage{graphicx}
\usepackage{pslatex}

\newif\ifprintedversion
\printedversiontrue

\ifprintedversion
\usepackage{hyperref}
\else
\usepackage{hyperref}
\fi

\makeatletter
\def\@maketitle{
  \global\setbox\fm@box=\vbox\bgroup
  \vskip -15mm
  \hfill{\parbox[t]{3.8cm}{\begin{flushright}DESY 00-129\\
     SPhT--T00/124\end{flushright}}}
    \vskip 8mm                    
    \raggedright                  
    \hyphenpenalty\@M             
    {\Large \@title \par}         
    \vskip\@bls                   
    {\normalsize                  
     \@author \par}               
    \vskip\@bls                   
    \@address                     
  \egroup
  \twocolumn[
    \unvbox\fm@box                
    \vskip\@bls                   
    \unvbox\abstract@box          
    \vskip 2pc]}                  
\makeatother

\def\msbar{{\ensuremath{\overline{{\rm MS}}}}}

\newcommand{\xb}{\bar x}

\newcommand{\as}{\alpha_s}
\newcommand{\eq}[1]{Eq.~(\ref{#1})}

\newcommand{\st}[1]{\mbox{\scriptsize #1}}
\newcommand{\nn}{\nonumber}

\newcommand{\e}{\epsilon}

\newcommand{\rg}{r_\Gamma}

\newcommand{\ycut}{y_{\st{cut}}}

\def\L{\left(}\def\R{\right)}

\def\Qb{{\overline{Q}}}

\def\xg{{x_g}}
\def\h6born{h_6^{\st{LO}} } 
 
\def\deltah6{\delta h_6}

\def\epem{{e^+e^-}} 
\def\Qb{{\ensuremath{\bar{Q}}}}

\title{Parity-violating 3-jet
observables for massive quarks to order $\alpha_s^2$
in $\epem$ annihilation 
}

\author{W. Bernreuther\address{Institut f\"ur Theoretische Physik,
RWTH Aachen, D-52056 Aachen, 
Germany}\thanks{{\tt breuther@physik.rwth-aachen.de},
research supported by BMBF, contract 05 HT9PAA 1}, 
A. Brandenburg\address{DESY Theory Group, D-22603 Hamburg, 
Germany}\thanks{{\tt arnd.brandenburg@desy.de}, research supported by
Heisenberg Fellowship of D.F.G.} 
and P. Uwer\address{Service de Physique Th{\'e}orique,
Centre d'Etudes de Saclay, 
\mbox{F-91191 Gif-sur-Yvette cedex}, 
France}\thanks{{\tt uwer@spht.saclay.cea.fr}, speaker at the conference}}

\begin{document}
\begin{abstract}
In this talk we discuss the calculation 
of the QCD corrections to parity-violating 
3-jet observables in $e^+e^-$ collisions, 
keeping the full quark mass dependence \cite{BeBrUw00}.
\end{abstract}

\thispagestyle{empty}
\maketitle

\section{Introduction}
The measurement of the forward-backward asymmetry $A_{\st{FB}}^b$ for 
$b$-quark production in
$e^+ e^-$ collisions at the Z peak provides one of the most precise
determinations of the weak mixing angle. 
It can be determined with an error at the per mille level if the 
computations of $A_{\st{FB}}^b$
are at least as accurate as the present experimental 
precision of about two percent \cite{LEP99}.
To achieve this precision one has to go beyond leading order in 
perturbation theory. Furthermore given the size of mass effects 
due to the non-zero $b$-quark mass, it is also necessary to extend existing
analyses to account for non-vanishing quark masses. This is mandatory if one
studies top-quarks instead of $b$-quarks.
Restricting ourselves to perturbative QCD the present knowledge is as follows.
The order $\alpha_s$ contributions had been computed first 
for massless \cite{JeLaZe81} and then for massive quarks 
\cite{DjKuZe90,ArBaLe92}.
These calculations, which used the quark direction for defining the asymmetry 
were later modified \cite{DjLaZe95,DjZe97,La96} to allow the use of the
thrust axis to define the asymmetry.
The next-to-next-to-leading order (NNLO) coefficient was computed first in 
ref. \cite{AlLa93}, for massless quarks using the quark axis definition.
This result was recently corrected 
by a completely analytical \cite{NeRa98} and by a numerical
calculation \cite{CaSe99b}. In ref. \cite{CaSe99b} the NNLO corrections
were also determined for the thrust axis definition.
To complete the analyses of $A_{\st{FB}}$ at NNLO the  
corrections for massive quarks are needed. 
To obtain these corrections it is convenient to compute the individual
contributions of the parton jets \cite{AlLa93}, in particular to
calculate separately the contributions from 2-, 3-, and 4-jet final states
involving a heavy quark $Q$.
For the 4-jet contribution to $A_{\st{FB}}$ one needs only the known 
Born matrix elements for $e^+e^- \to 4$ partons 
involving at least one ${\bar Q}Q$ pair (see, e.g. ref. \cite{BaMaMo94}).
The phase-space integration of these matrix elements
in order to get $(A_{\st{FB}})_{\st{4-jet}}$ for a given jet algorithm can
be done numerically.

Compared to $(A_{\st{FB}})_{\st{4-jet}}$ the computation 
of $(A_{\st{FB}})_{\st{3-jet}}$ is much more difficult. Here  the
NLO corrections to the partonic subprocess $\epem \to Q\Qb g$ are needed.
This contribution must be combined with 
the  3-jet contribution from subprocesses with four partons in the final state.
Only by combining the two contributions 
a result which is free of soft- and mass singularities is obtained.
In this talk we will present recent results \cite{BeBrUw00} 
on the missing ingredients
that are needed to calculate 
$(A_{\st{FB}})_{\st{3-jet}}$. 

The computation of $(A_{\st{FB}})_{\st{2-jet}}$ for heavy quarks 
is beyond the scope of this work and remains to be done in the future.

To fix our notation we discuss briefly in section \ref{sec:kinematics}  
the kinematics and the leading order results. 
In section \ref{sec:loop-corrections} we comment on the calculation 
of the one-loop corrections. In the following section \ref{sec:checks}
some checks
are described. We close with a conclusion in section \ref{sec:conclusions}.

\section{Kinematics and leading order results}
\label{sec:kinematics}
\def\bp{{\bf p}}
\def\bk{{\bf k}}
The leading order differential cross section for
\begin{equation}
  \label{eq:QQbg}
  e^+(p_+)e^-(p_-)\to Q(k_Q) \bar{Q}(k_{\bar{Q}})g(k_g)
\end{equation}
can be written in the following form{\footnote{
We neglect the lepton masses and do not consider transversely polarized
beams.}}: 
\begin{eqnarray}
\label{eq:dsigma}
  &&
  \frac{d\sigma}{d\phi d\!\cos\vartheta  dx d\xb }=
  \frac{3}{4}\frac{1}{(4\pi)^3}\sigma_{pt}\Bigg\{
  (1+\cos^2\vartheta)\, F_1\nn\\
  &+& (1-3\cos^2\vartheta)\,F_2 
  + \cos\vartheta\,F_3
  + \sin 2\vartheta \cos\phi\, F_4 \nonumber\\
  &+& \sin^2\vartheta \cos 2\phi\,F_5
  + \sin\vartheta\cos\phi\,F_6
  \Bigg\},
  \label{eq:diffcs}
\end{eqnarray}
with 
\begin{equation}
  \label{eq:sigmapt}
  \sigma_{pt} = \sigma(e^+e^-\rightarrow \gamma^\ast 
  \rightarrow\mu^+\mu^-) = 
  \frac{4\pi\alpha^2}{3s}.
\end{equation}
In \eq{eq:dsigma}, $\vartheta$ denotes the angle between the 
direction ${\bf p}_-$ of the incoming electron $e^-$
and the direction ${\bf k}_Q$ 
of the heavy quark $Q$. The angle $\phi$ is the oriented 
angle between the plane defined by $e^-$ and $Q$
 and the plane defined by the quark anti-quark pair.
The functions $F_i$ depend only on the scaled c.m. energies 
\begin{equation}
  x={2k\cdot k_Q\over s} \quad \mbox{and}\quad 
  \xb={2k\cdot k_{\bar{Q}}\over s},
\end{equation}
and on the scaled mass square 
\begin{equation}
  z={m^2\over s}
\end{equation}
of the heavy quark and anti-quark,
where $k=p_+ + p_-$ and $s=k^2$. Note that the leading order  
differential cross section for a final state with $n\ge 4$ 
partons can be written in an analogous 
way in terms of the two angles $\vartheta$,  $\phi$ and 
$3n-7$ variables that involve only the final state momenta.
In higher orders absorptive parts give rise to three additional 
functions $F_{7,8,9}$ (cf. ref. \cite{BrDiSh96}).
The four functions $F_{1,2,4,5}$ are parity even.
In particular, the 3-jet production rate is determined by $F_1$. 
In the case of massive quarks 
next-to-leading order results for $F_1$ were given
in references
\cite{BeBrUw97a,BrUw97,Ro96,Ro97,BiRoSa97b,BiRoSa99,NaOl97b,Ol97}.
The two functions $F_3$ and $F_6$ are induced by the interference
of a vector and an axial-vector current. In particular, 
using the quark-axis, the 3-jet 
forward-backward asymmetry is related to $F_3$ in leading order 
in the following way:
\begin{eqnarray}
\label{eq:defafb}
(A_{\st{FB}})_{\st{3-jet}} &=& 
{\int d\sigma\,\,   \theta_{\st{cut}}(\ycut) \,\,
{\rm sgn}(\cos(\vartheta))
\over 
{\int d\sigma\,\,   \theta_{\st{cut}}(\ycut) 
}}
\nonumber \\ 
&=& \frac{3}{8}
{\int dx d\xb \,\,\theta_{\st{cut}}(\ycut) \,\, F_3(x,\xb) 
  \over 
  \int dx d\xb \,\,\theta_{\st{cut}}(\ycut)\,\,  F_1(x,\xb) },
\end{eqnarray}    
where $\theta_{\st{cut}}(\ycut)$ defines, for a given jet finding algorithm 
and jet resolution parameter
$\ycut$, a region in the $(x,\xb)$ plane.
The electroweak couplings appearing in $F_{3,6}$ can be factored
out as follows:
\begin{eqnarray}
\label{eq:f36}
&&F_{3,6} =\nn\\
&&-g_a^Q(1-\lambda_+\lambda_-)\big[Q_Q {\rm Re} \chi (g_a^e
-f(\lambda_+,\lambda_-)g_v^e)\nn\\
&&+g_v^Q |\chi|^2 (f(\lambda_+,\lambda_-)(g_v^{e2}+g_a^{e2})-2g_v^eg_a^e)
 \big]\tilde{F}_{3,6}.
\end{eqnarray}
In \eq{eq:f36},
\begin{eqnarray}
g_v^f &=& T_3^f-2Q_f\sin^2\theta_W,\nonumber \\
g_a^f &=& T_3^f, \nonumber \\
\chi &=& {1\over 4\sin^2\theta_W\cos^2\theta_W}
{s\over s-m_Z^2+i m_Z\Gamma_Z}, \nonumber \\
f(\lambda_+,\lambda_-) &=& {\lambda_--\lambda_+\over 1-\lambda_-\lambda_+},
\end{eqnarray}
where $T_3^f$ is the third component of the weak isospin of the fermion $f$,
$\theta_W$ is the weak mixing angle, and  $\lambda_\mp$ denotes the 
longitudinal polarization of the electron (positron).
In next-to-leading order in $\alpha_s$, additional contributions to 
$F_{3,6}$ with electroweak couplings different from those 
in \eq{eq:f36} are induced which we
will not consider here. They are either proportional to ${\rm Im}\chi$
and thus suppressed formally in the electroweak coupling
or generated by the triangle fermion loop diagrams. 
The contribution of the triangle fermion loop
is gauge independent and UV and IR finite. It was calculated some time ago 
in ref. \cite{HaKuYa91}.

The functions $\tilde{F}_{3,6}(x,\xb)$ may be expressed 
in terms of  functions $h_6$, $h_7$ which appear in the 
decomposition of the so-called hadronic tensor 
as performed for example in references \cite{KoSc85,KoSc89,KoScKrLa86}.
\begin{eqnarray}
\label{eq:h6f36}
\tilde{F}_3 &=& {1\over 2}\big[\sqrt{x^2-4z}h_6(x,\xb)\nn\\
  &+&\sqrt{\xb^2-4z}\cos\vartheta_{Q\bar{Q}}h_7(x,\xb)\big],\nonumber \\
\tilde{F}_6 &=& -{1\over 2}\sqrt{\xb^2-4z}\sin\vartheta_{Q\bar{Q}}h_7(x,\xb),
\end{eqnarray}
where $\vartheta_{Q\bar{Q}}$ is the angle between $Q$ and $\bar{Q}$ in the
c.m. frame and we have
\begin{equation}
\cos\vartheta_{Q\bar{Q}}={2(1-x-\xb+2z)+x\xb
\over \sqrt{x^2-4z}\sqrt{\xb^2-4z}}.
\end{equation}
It would seem pointless to trade $F_{3,6}$ for $h_{6,7}$ if it were not
for the relation (which follows from CP invariance):
\begin{equation}
\label{eq:h7}
h_7(x,\xb)=-h_6(\xb,x).
\end{equation}  
To calculate the functions $F_{3,6}$ it is thus sufficient to determine
the function $h_6(x,\xb)$. In the one-loop corrections to $h_6$ one 
encounters both ultraviolet (UV) and infrared (IR) singularities.
Regulating the UV as well as the IR singularities by
continuation to $d=4-2\e$ space-time dimensions, we need the Born
result in $d$ dimensions:
\begin{eqnarray}
  \label{eq:bornresult}
 &&\h6born(x,\xb) 
 = 
 16\*\pi\*\as\*(N^2-1) \* B \nn\\
 &&\times\bigg(2\* x
 -(x\*\xb+\xb^2+2-4\*\xb)\*\e
 -4\*{\xg\over 1-x}\*z
 \bigg)
\end{eqnarray}
with 
\begin{equation}
  \label{eq:Bdef}
  B = {1\over (1-x)\*(1-\xb)}
  ,
\end{equation}
and the scaled gluon energy
\begin{equation}
  x_g = {2k\cdot k_g\over s}= 2-x-\xb.
\end{equation}
Note that the terms proportional to $\e$ in \eq{eq:bornresult} depend on the
prescription used to treat $\gamma_5$ in $d$ dimensions. To derive the
above equation we have used the prescription 
\begin{equation}
  \label{eq:g5def}
   \gamma_\mu \gamma_5\rightarrow 
  (1 -\frac{\as}{\pi} C_F){i\over3!}\varepsilon_{\mu\beta\gamma\delta}\,
  \gamma^\beta\gamma^\gamma\gamma^\delta
 \quad\cite{tHoVe72,La93}.
\end{equation}

\section{Virtual corrections}
\label{sec:loop-corrections}
Given the vast knowledge on how to perform one-loop calculations 
the derivation of the one-loop corrections to $h_6$ is in principle 
straightforward. So we restrict our discussion of the virtual
corrections to some technical remarks and the final results.

We used the background field gauge \cite{Ab81,AbGrSc83} with the 
gauge parameter set to one. In this gauge the three-gluon
vertex is simplified which leads to a reduction 
of the number of terms encountered
in intermediate steps of the calculation. Furthermore we used the 
Passarino-Veltman 
reduction procedure \cite{PaVe79} to reduce the one-loop tensor 
integrals to scalar one-loop integrals. As mentioned earlier
we regularized both
UV and IR singularities by  continuation to $d=4-2\e$ space-time 
dimensions. 
Note that the `t Hooft-Veltman prescription to 
treat $\gamma_5$ in $d$ dimensions
give rise to an additional `finite renormalization' 
(see \eq{eq:g5def}) to restore the chiral Ward identities.

The finite contributions of the loop-diagrams are too long to be 
reproduced here.
They are given explicitly in ref. \cite{BeBrUw00}. Here we give only
the results for the UV and IR divergences and the contribution from
the renormalization procedure. In the following we denote with $m$
always the mass parameter renormalized in the on-shell scheme and with
$\as$ the strong coupling in the usual $\msbar$ scheme evaluated 
at the renormalization scale $\mu$. 

Keeping only the pole-part of the integrals we obtain the following 
result for the UV singularities:
\begin{eqnarray}
  \label{eq:UVsingular}
  h_6^{\st{virt., UV div.}}
  &=& {\rg {1\over \e} \L{m^2\over 4\pi\mu^2}\R^{-\e}}
  \bigg\{
  {\as\over 2\*\pi} \*C_F\* \h6born\nn\\
  &&\hspace{-2.5cm}
  +\,\,24 \*\as^2\*(N^2-1)\*C_F \* \deltah6
  +8\*\as^2\* (N^2-1)\*C_F\* B\*\e\*\bigg( 
  1+x\nn\\
  &&\hspace{-1.5cm}-\,\,2\*\xb+(1-\xb)\*{\xb\over x}+z\*\bigg[2\*B\*
  ((\xb^2-2\*x)\*(1-x)
  \nn\\
  &&
  \hspace{-1.5cm}+\,\,(2-3\*x)\*(1-\xb))
  +g_1\*{x^2-4\*z\over x\*(1-x)}\nn\\
  &&
  \hspace{-1.5cm}+\,\,4\*z\*B\*(2\*{(1-\xb)^2\over 1-x}+\xg)\bigg]
  \bigg)\nn\\
  &&\hspace{-1.5cm}+ 16\*\as^2\*C_F \*B\*\e\*(
  {1\over \xg}\*(x\*\xb-4\*\xb+2+\xb^2)\nn\\
  &&-\,\,2\*z\*\xg\*B\*(x-\xb)
  ) + O(\e^2)\bigg\}
\end{eqnarray}
with
\begin{eqnarray}
  \deltah6 &=& B^2\*
  \bigg(\bigg[-2\*(x^2-3\*x-\xb^2\*x+3\*x\*\xb+2\*\xb^2\nn\\
   &-& 4\*\xb+2)
  -\xg\*(\xb^2+x\*\xb-4\*\xb+2)\*\e
  \bigg]\*z\nn\\
  &&\hspace{-1.cm}-\,\,
  4\*{1\over 1-x}\*(-5\*\xb-3\*x+2\*\xb^2+4+x^2+x\*\xb)\*z^2
  \bigg),
\end{eqnarray}
and
\begin{equation}
  \label{eq:g_1def}
  g_1 = -{(1-x)\*(x-2\*\xb)\over (x^2-4\*z)^2}
  \*(x\*(\xb+x)-2\*(1-\xg))
  .
\end{equation}
The usual one-loop factor $\rg$ is given by:
\begin{equation}
  \rg = {\Gamma(1+\e)\*\Gamma^2(1-\e)\over \Gamma(1-2\e)}.
\end{equation}
The IR divergent contributions from the loop-integrals are given by
\begin{eqnarray}
  h_6^{\st{virt., IR div.}}
  &=&
  {\as\over 2\*\pi} \* \L {4\*\pi\*\mu^2  \over m^2 } \R ^{\e}
  \*\rg \*\h6born\*\bigg\{\nn\\
  &-&
  {N}\* 
  \left[
    {1\over \e^2}
    - {1\over \e}\* \ln({t_{Qg}\over m^2})
    - {1\over \e}\* \ln({t_{\bar Qg}\over m^2})
  \right]\nn\\
  &+&{1\over N}
  \* {1\over \e}
\*{1+\omega^2\over 1-\omega^2}\* \ln(\omega)\bigg\},
\end{eqnarray}
with $t_{ij} = 2 k_i\cdot k_j$ and 
\begin{equation}
  \omega =   {1-\sqrt{1-{4\*z\over x+\xb-1}}
    \over 1+\sqrt{1-{4\*z\over x+\xb-1}}}.
\end{equation}
Finally, the entire contribution from renormalization 
(including the Lehmann--Symanzik--Zimmermann residue and the 
`$\gamma_5$-counterterm') is given by:
\begin{eqnarray}
  h_6^{\st{virt., ren.}}
  &=&
  -{\as\over 2\*\pi}\* \rg\*C_F 
  \L{m^2\over 4\pi\mu^2}\R ^{-\e} \* 
   {1\over \e}\h6born\nn\\
  &&\hspace{-2cm}- 24\as^2   (N^2-1) C_F
  \rg \L {m^2\over 4\pi\mu^2}\R ^{-\e} {1\over \e} 
  \*
  \deltah6\nn\\
  &&\hspace{-2cm}-{\as\over 4\*\pi}\*\rg\*  \L{m^2\over 4\pi\mu^2}\R ^{-\e} 
  {1\over \e}\left\{ 4  C_F 
  -  
  \*\left({2\over 3}\* n_f^l -{11\over 3}\* N\right) 
  \right\}\h6born \nn\\
  &&\hspace{-2.2cm}+{\as\over 4\*\pi}\*  
  \left\{ ({2\over 3}\* n_f -{11\over 3}\* N)\ln({m^2\over \mu^2})
  +{2\over 3}\*\sum_i \ln{m_i^2\over m^2} - 8  C_F\right\}  \h6born\nn\\
  &&\hspace{-2.cm}- 32\as^2   (N^2-1) C_F
  \deltah6
  -\frac{\as}{\pi} C_F \h6born
  \label{eq:ren-contr}
\end{eqnarray}
Note that the singularity in the third line of \eq{eq:ren-contr}
is a collinear singularity and not a UV singularity.

\section{Checks}
\label{sec:checks}
In this section we discuss some checks to assure the correctness of 
our results. Given the fact that the finite results are given in the
form (coefficients)$\times$(one-loop integrals) and that the loop integrals
are known, it is sufficient to check the  coefficients.
As one can see from the comparison of \eq{eq:UVsingular} and
\eq{eq:ren-contr} the UV divergences from the loop diagrams
are cancelled exactly by the UV singularities from the renormalization 
procedure. This
cancellation is thus an excellent check of the coefficients of the UV 
divergent integrals, namely the one- and two-point integrals. 
We checked also that the finite renormalization due to the 
`t~Hooft-Veltman prescription restores the chiral Ward identities.

According
to the Kinoshita-Lee-Nauenberg theorem \cite{LeNa64,Ki65} the remaining
IR divergences must be cancelled by the real corrections.
To check this cancellation we have calculated the singular contributions
from real emission by using a modified version \cite{BrUw97,KeLa99} of 
the phase space slicing method \cite{GiGl92,GiGlKo93}.
By this cancellation  the coefficients of the IR divergent
triangle integrals and the box integrals are checked. The explicit
results for the singular contributions of the real corrections
are given in ref. \cite{BeBrUw00}.

Furthermore we have verified that we are able to reproduce the coefficients
of the loop-integrals in the massless result \cite{KoSc85,KoSc89,KoScKrLa86}.
We found agreement with Eq.~(4.10) of ref. \cite{KoSc89}. 
In the case of  Eq.~(4.9) of ref. \cite{KoSc89} we found agreement 
only up to an overall factor $(-1)$. 
This is most probably caused by  a typographical error  
which is also present in ref. \cite{KoSc85},  but not in 
ref. \cite{KoScKrLa86} with which we fully agree.

\section{Conclusions}
\label{sec:conclusions}
In this talk we have presented the necessary ingredients for 
calculating virtual corrections to parity-violating three jet
observables. Together with the existing results on the real contribution 
\cite{BaMaMo94}, in particular the 
singular contributions \cite{BeBrUw00},
these results allow for the
calculation of parity-violating three-jet
observables to next-to-leading order accuracy.

\ifprintedversion
\bibliographystyle{h-elsevier2}
\else
\bibliographystyle{utcaps}
\fi

\bibliography{literatur}

\newcommand{\zp}{Z. Phys. }\def\as{\alpha_s }\newcommand{\prd}{Phys. Rev.
  }\newcommand{\pr}{Phys. Rev. }\newcommand{\prl}{Phys. Rev. Lett.
  }\newcommand{\npb}{Nucl. Phys. }\newcommand{\psnp}{Nucl. Phys. B (Proc.
  Suppl.) }\newcommand{\pl}{Phys. Lett. }\newcommand{\ap}{Ann. Phys.
  }\newcommand{\cmp}{Commun. Math. Phys. }\newcommand{\prep}{Phys. Rep.
  }\newcommand{\jmp}{J. Math. Phys. }\newcommand{\rmp}{Rev. Mod. Phys. }
\begin{thebibliography}{10}

\bibitem{BeBrUw00}
W. Bernreuther, A. Brandenburg and P. Uwer,
\newblock Next-to-leading order {QCD} corrections to parity-violating 3-jet
  observables for massive quarks in $e^+e^-$ annihilation, 2000,
  \href{http://www.arXiv.org/abs/hep-ph/0008291}{{\tt hep-ph/0008291}}.

\bibitem{LEP99}
D. Abbaneo et~al.,
\newblock A combination of preliminary electroweak measurements and constraints
  on the standard model, 1999,
\newblock {CERN-EP-99-015}.

\bibitem{JeLaZe81}
J. Jersak, E. Laermann and P.M. Zerwas,
\newblock \prd D25 (1982) 1218.

\bibitem{DjKuZe90}
A. Djouadi, J.H. K{\"u}hn and P.M. Zerwas,
\newblock \zp C46 (1990) 411.

\bibitem{ArBaLe92}
A.B. Arbuzov, D.Y. Bardin and A. Leike,
\newblock Mod. Phys. Lett. A7 (1992) 2029.

\bibitem{DjLaZe95}
A. Djouadi, B. Lampe and P.M. Zerwas,
\newblock \zp C67 (1995) 123,
  \href{http://www.arXiv.org/abs/hep-ph/9411386}{{\tt hep-ph/9411386}}.

\bibitem{DjZe97}
A. Djouadi and P.M. Zerwas,
\newblock Memorandum: {QCD} corrections to {A(FB)}(b),
\newblock {DESY-T-97-04}.

\bibitem{La96}
B. Lampe,
\newblock A note on {QCD} corrections to {A(b)(FB)} using thrust to determine
  the b-quark direction, 1996,
  \href{http://www.arXiv.org/abs/hep-ph/9812492}{{\tt hep-ph/9812492}}.

\bibitem{AlLa93}
G. Altarelli and B. Lampe,
\newblock Nucl. Phys. B391 (1993) 3,
\newblock 

\bibitem{NeRa98}
V. Ravindran and W.L. van Neerven,
\newblock Phys. Lett. B445 (1998) 214,
  \href{http://www.arXiv.org/abs/hep-ph/9809411}{{\tt hep-ph/9809411}},
\newblock 

\bibitem{CaSe99b}
S. Catani and M.H. Seymour,
\newblock JHEP 07 (1999) 023,
  \href{http://www.arXiv.org/abs/hep-ph/9905424}{{\tt hep-ph/9905424}},
\newblock 

\bibitem{BaMaMo94}
A. Ballestrero, E. Maina and S. Moretti,
\newblock \npb B415 (1994) 265,
  \href{http://www.arXiv.org/abs/hep-ph/9212246}{{\tt hep-ph/9212246}}.

\bibitem{BrDiSh96}
A. Brandenburg, L. Dixon and Y. Shadmi,
\newblock \prd D53 (1996) 1264,
  \href{http://www.arXiv.org/abs/hep-ph/9505355}{{\tt hep-ph/9505355}}.

\bibitem{BeBrUw97a}
W. Bernreuther, A. Brandenburg and P. Uwer,
\newblock \prl 79 (1997) 189,
  \href{http://www.arXiv.org/abs/hep-ph/9703305}{{\tt hep-ph/9703305}}.

\bibitem{BrUw97}
A. Brandenburg and P. Uwer,
\newblock \npb B515 (1998) 279,
  \href{http://www.arXiv.org/abs/hep-ph/9708350}{{\tt hep-ph/9708350}}.

\bibitem{Ro96}
G. Rodrigo,
\newblock Quark mass effects in {QCD} jets,
\newblock PhD thesis, University of Valencia, 1996,
  \href{http://www.arXiv.org/abs/hep-ph/9703359}{{\tt hep-ph/9703359}}.

\bibitem{Ro97}
G. Rodrigo,
\newblock \psnp 54A (1997) 60,
  \href{http://www.arXiv.org/abs/hep-ph/9609213}{{\tt hep-ph/9609213}}.

\bibitem{BiRoSa97b}
G. Rodrigo, A. Santamaria and M. Bilenky,
\newblock \prl 79 (1997) 193,
  \href{http://www.arXiv.org/abs/hep-ph/9703358}{{\tt hep-ph/9703358}}.

\bibitem{BiRoSa99}
G. Rodrigo, M. Bilenky and A. Santamaria,
\newblock Nucl. Phys. B554 (1999) 257,
  \href{http://www.arXiv.org/abs/hep-ph/9905276}{{\tt hep-ph/9905276}},
\newblock 

\bibitem{NaOl97b}
P. Nason and C. Oleari,
\newblock \npb B521 (1998) 237,
  \href{http://www.arXiv.org/abs/hep-ph/9709360}{{\tt hep-ph/9709360}}.

\bibitem{Ol97}
C. Oleari,
\newblock Next-to-leading-order corrections to the production of heavy-flavour
  jets in e+ e- collisions,
\newblock PhD thesis, Milan U., 1997,
  \href{http://www.arXiv.org/abs/hep-ph/9802431}{{\tt hep-ph/9802431}},
\newblock 

\bibitem{HaKuYa91}
K. Hagiwara, T. Kuruma and Y. Yamada,
\newblock \npb B358 (1991) 80.

\bibitem{KoSc85}
J.G. K{\"o}rner and G. Schuler,
\newblock \zp 26 (1985) 559.

\bibitem{KoSc89}
G.A. Schuler and J.G. K{\"o}rner,
\newblock Nucl. Phys. B325 (1989) 557,
\newblock 

\bibitem{KoScKrLa86}
J.G. K{\"o}rner et~al.,
\newblock \zp 32 (1986) 181.

\bibitem{tHoVe72}
G. 't~Hooft and M. Veltman,
\newblock \npb B44 (1972) 189.

\bibitem{La93}
S.A. Larin,
\newblock \pl B303 (1993) 113.

\bibitem{Ab81}
L.F. Abbott,
\newblock \npb B185 (1981) 189.

\bibitem{AbGrSc83}
L.F. Abbott, M.T. Grisaru and R.K. Schaefer,
\newblock \npb B229 (1983) 372.

\bibitem{PaVe79}
G. Passarino and M. Veltman,
\newblock \npb B160 (1979) 151.

\bibitem{LeNa64}
T.D. Lee and M. Nauenberg,
\newblock \pr 133B (1964) 1549.

\bibitem{Ki65}
T. Kinoshita,
\newblock \jmp 3 (1965) 650.

\bibitem{KeLa99}
S. Keller and E. Laenen,
\newblock Phys. Rev. D59 (1999) 114004,
  \href{http://www.arXiv.org/abs/hep-ph/9812415}{{\tt hep-ph/9812415}},
\newblock 

\bibitem{GiGl92}
W.T. Giele and E.W.N. Glover,
\newblock \prd D46 (1992) 1980.

\bibitem{GiGlKo93}
W.T. Giele, E.W.N. Glover and D.A. Kosower,
\newblock \npb B403 (1993) 633,
  \href{http://www.arXiv.org/abs/hep-ph/9302225}{{\tt hep-ph/9302225}}.

\end{thebibliography}
\end{document}